\documentclass[12pt]{article}

\newcommand{\lesssim}{\:\mbox{\raisebox{-3pt}{$\stackrel%
{\displaystyle <}{\sim}$}}\:}
\newcommand{\gtrsim}{\:\mbox{\raisebox{-3pt}{$\stackrel%
{\displaystyle >}{\sim}$}}\:}

\begin{document}

\title{\normalsize \hfill UWThPh-2000-26 \\[1cm] \LARGE
A neutrino mass matrix with seesaw mechanism \\
and two-loop mass splitting}

\author{
W.\ Grimus \\
\small Universit\"at Wien, Institut f\"ur Theoretische Physik \\
\small Boltzmanngasse 5, A--1090 Wien, Austria
\\[3mm]
L.\ Lavoura \\
\small Universidade T\'ecnica de Lisboa \\
\small Centro de F\'\i sica das Interac\c c\~oes Fundamentais \\
\small Instituto Superior T\'ecnico, 1049-001 Lisboa, Portugal
}

\date{3 July 2000}

\maketitle

\begin{abstract}
We propose a model which uses the seesaw mechanism
and the lepton number $\bar L = L_e - L_\mu - L_\tau$
to achieve the neutrino mass spectrum
$m_1 = m_2$ and $m_3 = 0$,
together with a lepton mixing matrix $U$ with $U_{e3} = 0$. 
In this way,
we accommodate atmospheric neutrino oscillations.
A small mass splitting $m_1 > m_2$
is generated by breaking $\bar L$ spontaneously
and using Babu's two-loop mechanism.
This allows us to incorporate ``just so'' solar-neutrino oscillations
with maximal mixing into the model.
The resulting mass matrix has three parameters only,
since $\bar L$ breaking
leads exclusively to
a non-zero $ee$ matrix element.  
\end{abstract}

\vspace{6mm}

The recent results of Super-Kamiokande \cite{SK-atm},
providing evidence for atmospheric neutrino oscillations,
have lead to increased efforts \cite{barr}
to investigate mechanisms for the generation of neutrino masses and mixings.
In this context one wants to solve two questions:
\begin{enumerate}
\item Why are the neutrino masses so much smaller
than the charged-lepton masses?
\item How can the specific features of the neutrino mass spectrum
and of the
lepton
mixing matrix \cite{review},
needed to reproduce the atmospheric and solar-neutrino deficits,
be generated within a model?
\end{enumerate}
Proposals to answer the first question
are given by the seesaw mechanism \cite{seesaw}
and by radiative neutrino-mass generation.
If we confine ourselves to extensions of the Standard Model
in the Higgs sector \cite{konetschny},
then purely radiative neutrino masses are obtained
within the models of Zee \cite{zee,jarlskog}
and of Babu \cite{babu},
and extended versions thereof \cite{joshi,luis,kitabayashi}.
As for the second question,
one prominent feature of the mixing matrix $U$
is the smallness of $U_{e3}$ \cite{CHOOZ,Ue3},
for which one
would like
to find an explanation in a model.
As a means to achieve this,
the lepton number $\bar L = L_e - L_\mu - L_\tau$
has been suggested \cite{barbieri}.

In this letter
we propose a model which combines the seesaw mechanism,
Babu's radiative two-loop mechanism,
and the lepton number $\bar L$.
In this way,
we have an explanation for the smallness of the neutrino masses.
The seesaw mechanism will enable us to
fit
the atmospheric neutrino oscillations,
whereas Babu's mechanism will generate
the
small mass-squared difference necessary for solar-neutrino oscillations.
The lepton number $\bar L$ will insure $U_{e3} = 0$.
However,
the breaking of this lepton number
is crucial for solar-neutrino oscillations.

Our model is given by the Standard Model of electroweak interactions,
based on the gauge group SU(2)$\times$U(1),
with three Higgs doublets
$\phi_k = \left( \varphi_k^+, \varphi_k^0 \right)^T$,
where $k = 1, 2, 3$.
The vacuum expectation values of 
$\sqrt{2} \varphi^0_k$ are denoted $v_k$.
We also introduce two neutrino singlets $\nu_{Rj}$ ($j=1,2$)
and the scalar singlets $f^+$, $h^{++}$, and $\eta$.
The latter scalar is complex but has zero electric charge.
We have the following assignments of the lepton number $\bar L$
to these multiplets:
\begin{equation}\label{Lbar}
\renewcommand{\arraystretch}{1.3}
\begin{array}{c|ccccccccccc}
& \nu_e, e & \nu_\mu, \mu & \nu_\tau, \tau & \nu_{R1} & \nu_{R2} &
\phi_1 & \phi_2 & \phi_3 & f^+ & h^{++} & \eta \\ \hline
\bar L & 1 & -1 & -1 & 1 & -1 & 0 & 0 & 0 & 0 & 2 & -2
\end{array}\ .
\end{equation}
Furthermore,
we need a discrete symmetry $S$ defined by
\begin{equation}\label{S}
\begin{array}{cccccc}
S: & \ell_R \to i \ell_R \quad (\ell = e,\mu,\tau), &
\phi_2 \to -i \phi_2, & \phi_3 \to -\phi_3, & h^{++} \to -h^{++}, 
& \eta \to -\eta,
\end{array}
\end{equation}
while
all other multiplets,
in particular the left-handed lepton doublets,
transform trivially under $S$.
With Eqs.~(\ref{Lbar}) and (\ref{S}) we obtain the Yukawa Lagangian
\begin{eqnarray}
{\cal L}_{\rm Y} &=& 
-\frac{\sqrt{2}}{v_2}
\sum_{\ell = e,\mu,\tau} 
m_\ell\, (\bar\nu_{\ell L},\bar\ell_L) 
\left( \begin{array}{c} \varphi^+_2 \\ \varphi^0_2 \end{array} \right)
\ell_R 
\nonumber\\[1mm]
 & &
- \frac{\sqrt{2}}{v_1^\ast}
\left[
\delta_e^\ast\, (\bar\nu_{eL},\bar e_L) \nu_{R1}
+ \delta_\mu^\ast\, (\bar\nu_{\mu L},\bar\mu_L) \nu_{R2}
+ \delta_\tau^\ast\, (\bar\nu_{\tau L},\bar\tau_L) \nu_{R2}
\right]
\left( \begin{array}{c} {\varphi^0_1}^\ast \\ - \varphi^-_1
\end{array} \right)
\nonumber\\[1mm]
 & &
+ f^+
\left[
f_\mu \left(
\nu_{eL}^T C^{-1} \mu_L - e_L^T C^{-1} \nu_{\mu L}
\right)
+
f_\tau \left(
\nu_{eL}^T C^{-1} \tau_L - e_L^T C^{-1} \nu_{\tau L}
\right)
\right]
\nonumber\\[2mm]
 & &
+ \sum_{\ell, \ell' = \mu, \tau} 
h_{\ell \ell'}\, \ell_R^T C^{-1} \ell'_R h^{++}
+ {\rm H.c.},
\label{yukawa}
\end{eqnarray}
where $\delta_\alpha$ ($\alpha = e,\mu,\tau$),
$f_\mu$,
$f_\tau$,
and the $h_{\ell \ell'}$ are complex coupling constants.
The Yukawa Lagrangian in Eq.~(\ref{yukawa}) is the most general one
built out of the multiplets in our model
and compatible with the gauge symmetry
and with the symmetries $\bar L$ and $S$.
The first line of Eq.~(\ref{yukawa})
displays the ordinary Yukawa couplings to the Higgs doublet $\phi_2$,
which give mass to the charged leptons.
Notice that we have taken,
without loss of generality,
those Yukawa couplings to be
flavor-diagonal.
The second line of Eq.~(\ref{yukawa}) shows the Yukawa couplings
of the right-handed neutrino singlets to $\phi_1$;
the third and fourth line display the Yukawa couplings
needed to implement radiative neutrino masses
using Babu's mechanism \cite{babu}.
Notice that the symmetry $S$ forbids
the couplings $\nu_{R1}^T C^{-1} \nu_{R1} \eta$
and $\nu_{R2}^T C^{-1} \nu_{R2} \eta^*$,
which would be allowed by $\bar L$.

Since the right-handed neutrino fields are
gauge singlets,
we can write down the mass term 
\begin{equation}\label{LM}
{\cal L}_M = M^\ast\, \nu_{R1}^T C^{-1} \nu_{R2}
+ {\rm H.c.},
\end{equation}
which is compatible with the lepton number $\bar L$.
In the following,
$M$ will play the role of a large seesaw scale.

The neutral complex scalar $\eta$
breaks the lepton number $\bar L$ spontaneously
through its vacuum expectation value $\langle \eta \rangle_0$.
In this way,
the only term in the Higgs potential which is linear in $\eta$,
\begin{equation}
V_\eta = \lambda \eta f^- f^- h^{++} + {\rm H.c.},
\end{equation}
where $\lambda$ is a dimensionless coupling constant,
transforms upon spontaneous symmetry breaking into
\begin{equation}\label{sb}
V_{\rm sb} = \lambda \langle \eta \rangle_0 f^- f^- h^{++} + {\rm H.c.}
\end{equation}
$V_{\rm sb}$ provides the trilinear scalar coupling
required by Babu's mechanism \cite{babu}.
Note that one cannot introduce {\it a priori}
the $\bar L$-soft-breaking term $f^- f^- h^{++}$,
since this term has dimension 3
and then we would also have to introduce the mass terms
$\nu_{Rj}^T C^{-1} \nu_{Rj}$ ($j=1,2$),
which also break $\bar L$ softly
and have dimension 3;
else we would not have a technically natural model \cite{hooft}.
These mass terms would destroy the $\bar L$ invariance
of the light-neutrino mass matrix already at tree level.

The symmetry $S$ necessitates the introduction
of two Higgs doublets
in order to have enough freedom to make the charged leptons massive.
The Higgs doublet $\phi_3$
does not couple to the leptons due to the symmetry $S$,
but it is needed in order to have the terms
$( \phi_1^\dagger \phi_3 )^2$
and $( \phi_1^\dagger \phi_2 )
( \phi_3^\dagger \phi_2 )$ in the Higgs potential.
These terms prevent the appearance of a Goldstone boson
which would couple directly to the leptons.
On the other hand,
the spontaneous breaking of $\bar L$ results in a Majoron,
given,
if we assume that $\langle \eta \rangle_0$ is real,
by $\sqrt{2}\, \mbox{Im}\, \eta$.
However,
this Majoron is only very weakly coupled to matter
(electrons, up quarks, and down quarks)
via loop diagrams \cite{majoron}.

Notice that,
due to the specific form of the symmetry $S$,
there are no couplings of the type $f^- \phi_k^T \tau_2
\phi_{k^\prime}$,
where $\tau_2$ is the antisymmetric Pauli matrix
and therefore $k \neq k^\prime$.
The absence of these couplings impedes Zee's mechanism
for one-loop neutrino masses \cite{zee},
contrary to what happens in other models \cite{joshi}.

Let us now discuss the neutrino mass matrix.
We have a $5 \times 5$ Majorana mass matrix ${\cal M}$
following the five chiral neutrino fields in our model.
The neutrino mass term is given by
\begin{equation}
{\cal L}_{\rm D+M} =
\frac{1}{2}\, \Omega_L^T C^{-1} {\cal M} \Omega_L + {\rm H.c.},
\quad \mbox{with} \quad
\Omega_L =
\left( \nu_{eL}, \nu_{\mu L}, \nu_{\tau L},
(\nu_{R1})^c, (\nu_{R2})^c \right)^T.
\end{equation}
The mass matrix has the decomposition
\begin{equation}
{\cal M} = \left(
\begin{array}{cc} M_L & M_D^T \\ M_D & M_R \end{array}
\right),
\end{equation}
with, at tree level,
\begin{equation}
M_D =
\left( 
\begin{array}{ccc} \delta_e & 0 & 0 \\ 0 & \delta_\mu & \delta_\tau
\end{array}
\right)
\quad \mbox{and} \quad
M_R = \left( \begin{array}{cc} 0 & M \\ M & 0 \end{array}
\right),
\end{equation}
according to the second line of Eq.~(\ref{yukawa})
and to Eq.~(\ref{LM}).
The matrix $M_L$ vanishes at tree level.
Thus,
at tree level we have \cite{branco}
two degenerate neutrinos with large masses,
two degenerate neutrinos which are light due to the seesaw mechanism,
and one massless Weyl neutrino.
But,
since the lepton number $\bar L$ is broken
by $V_{\rm sb}$ in Eq.~(\ref{sb}),
$M_L \neq 0$ is generated at two loops by Babu's mechanism.
However,
as $\bar L$ is operative in the Yukawa couplings of $f^+$ and of $h^{++}$, 
one finds that only $(M_L)_{ee}$ is non-zero.
We obtain
\begin{equation}\label{2-loop}
M_L = \left( 
\begin{array}{ccc} m_{ee} & 0 & 0 \\ 0 & 0 & 0 \\ 0 & 0 & 0 \end{array}
\right),
\quad \mbox{with} \quad
m_{ee} = - \frac{2\lambda \langle \eta \rangle_0 }{(16 \pi^2)^2}\, 
\sum_{\ell, \ell' = \mu, \tau}
f_\ell h_{\ell \ell'} f_{\ell'} m_\ell m_{\ell'} I_{\ell \ell'}.
\end{equation}
Here,
$I_{\ell \ell'}$ is a convergent two-loop integral.
Assuming that the masses of $f^+$ and of $h^{++}$
are of the same order of magnitude
and are much larger than the masses of the charged leptons,
one finds \cite{babu,luis} $I_{\ell \ell'} \sim 1/m_h^2$,
since all (di-)logarithms are of order 1.

We may now apply the seesaw formula
in order to obtain the neutrino Majorana mass term
for the light neutrinos as
\begin{equation}
{\cal L}_{\rm m} = 
\frac{1}{2} \omega_L^T C^{-1} M_\nu\, \omega_L + {\rm H.c.},
\quad \mbox{with} \quad
\omega_L = \left( \nu_{eL}, \nu_{\mu L}, \nu_{\tau L} \right)^T,
\end{equation}
where
\begin{equation}\label{seesaw}
M_\nu = M_L - M_D^T M_R^{-1} M_D.
\end{equation}
Inserting the expressions for $M_L$, $M_D$, and $M_R$
into Eq.~(\ref{seesaw}),
we arrive at
\begin{equation}\label{Mnu}
M_\nu = \left(
\begin{array}{ccc}
m_{ee} & - \delta_e \delta_\mu/M &  - \delta_e \delta_\tau/M \\
- \delta_e \delta_\mu/M & 0 & 0 \\
- \delta_e \delta_\tau/M & 0 & 0
\end{array}
\right).
\end{equation}
Obviously,
the second term on the right-hand side of Eq.~(\ref{seesaw}),
{\it i.e.},
Eq.~(\ref{Mnu}) with $m_{ee} = 0$,
is invariant under the lepton number $\bar L$. 

By phase transformations,
all elements of the mass matrix in Eq.~(\ref{Mnu})
can be made real and non-negative.
Consequently,
there is no CP violation associated with this neutrino mass matrix.
The matrix is exceedingly simple,
since it is parametrized by only three positive quantities $a$, $r$,
and $b$:\footnote{As a matter of fact,
the mass matrix in Eq.~(\ref{M'})
is the one implicitly suggested by Joshipura and Rindani \cite{joshi}
at the end of their paper.
However,
these authors relied on Zee's mechanism instead of relying
on the seesaw mechanism,
and consequently they ran into difficulties with the orders of magnitude
of the various terms needed in order to fit both
the atmospheric and the solar-neutrino oscillations.
They have, therefore, discarded it.}
\begin{equation}\label{M'}
M'_\nu = \left(
\begin{array}{ccc}
a & rb & b \\ rb & 0 & 0 \\ b & 0 & 0
\end{array}
\right).
\end{equation}
The eigenvalues of $M'_\nu$,
expressed as neutrino masses,
are $m_1$, $-m_2$, and $m_3 = 0$, with
\begin{equation}\label{mass}
m_1 = \sqrt{m_0^2 + \frac{a^2}{4}} + \frac{a}{2} 
\quad \mbox{and} \quad
m_2 = \sqrt{m_0^2 + \frac{a^2}{4}} - \frac{a}{2} \,,
\end{equation}
where we have defined
\begin{equation}
m_0 \equiv b \sqrt{X}, \quad \mbox{with} \quad X \equiv 1+r^2 \,.
\end{equation}
Furthermore,
from $M'_\nu$ we derive the lepton mixing matrix
\begin{equation}\label{U}
U = \left(
\begin{array}{ccc}
c & -is & 0 \\ 
\frac{r}{\sqrt{X}}s & i\frac{r}{\sqrt{X}}c & \frac{1}{\sqrt{X}} \\[0.5mm]
\frac{1}{\sqrt{X}}s & i\frac{1}{\sqrt{X}}c & -\frac{r}{\sqrt{X}} 
\end{array}
\right)
\quad \mbox{with} \quad
c \equiv \cos \theta_\odot, \: s \equiv \sin \theta_\odot, 
\end{equation}
where $\theta_\odot$ is the solar-neutrino mixing angle.
Notice that $U_{e3} = 0$.
This is an important and exact prediction of our model.

Since the matrix element $a$ in $M'_\nu$
is meant to generate a small mass splitting between $m_1$ and $m_2$,
necessary in order to accommodate solar-neutrino oscillations,
this element must be tiny.
Therefore,
using the neutrino masses in Eq.~(\ref{mass}) together with $m_3 = 0$,
we find
\begin{equation}\label{m2}
\Delta m^2_{\rm atm} \simeq m_0^2 
\quad \mbox{and} \quad
\Delta m^2_\odot \simeq 2 m_0 a
\end{equation}
for the atmospheric and solar mass-squared differences,
respectively.
This in turn gives
\begin{equation}\label{adet}
a \simeq \frac{\Delta m^2_\odot}{2 \sqrt{\Delta m^2_{\rm atm}}}
\quad \mbox{and} \quad
\sin^2 2\theta_\odot \simeq 1 - 
\frac{1}{16} \left( \frac{\Delta m^2_\odot}{\Delta m^2_{\rm atm}} \right)^2.
\end{equation}
As a consequence,
the solar-neutrino mixing angle is
for all practical purposes $45^\circ$,
and a fit to the solar-neutrino data
requires the ``just so'' or vacuum-oscillation (VO) solution
to the solar-neutrino deficit
(for recent analyses see Refs.~\cite{VO,barger,goswami,quasi}).
The so-called LOW solution \cite{VO,quasi,GG}
might also be allowed \cite{dark}.
In the following we shall however concentrate on the VO solution.
Due to $U_{e3} = 0$, solar-neutrino oscillations are decoupled
from the atmospheric oscillation parameters \cite{giunti}.
The mixing angle for atmospheric neutrino oscillations
is obtained from the mixing matrix in Eq.~(\ref{U}) as
\begin{equation}\label{rdet}
\sin^2 2\theta_{\rm atm} = \frac{4r^2}{(1+r^2)^2}.
\end{equation}
According to the Super-Kamiokande results \cite{sobel},
$\sin^2 2\theta_{\rm atm} \gtrsim 0.88$ at 90\% CL,
and one therefore has the range
\begin{equation}\label{rvalue}
0.7 \lesssim r \lesssim 1.4
\end{equation}
for the parameter $r$.  
We then have a scenario very close to bimaximal mixing \cite{bimaximal}.

Thus,
with Eqs.~(\ref{adet}), (\ref{rvalue}), and
\begin{equation}
b \simeq \sqrt{\frac{\Delta m^2_{\rm atm}}{1+r^2}},
\end{equation}
all three parameters of $M'_\nu$ are in principle determined by
the atmospheric and solar-neutrino oscillation data.
Summarizing,
with $\Delta m^2_\odot \sim 10^{-10}$ eV$^2$,
$\Delta m^2_{\rm atm} \sim 3.2 \times 10^{-3}$ eV$^2$ \cite{sobel},
and the atmospheric neutrino mixing angle close to $45^\circ$,
we arrive at the order-of-magnitude estimates
\begin{equation}\label{estimate}
a \sim 10^{-9} \; \mbox{eV}, \quad
b \sim 0.04 \; \mbox{eV}, \quad r \sim 1.
\end{equation}

We now want to study how to implement these estimates in our model.
To this end
it is useful to consider the different mass scales present in the model.
Apart from the electroweak scale,
and with the following simplifying assumptions,
we have three scales:
\begin{itemize}
\item The scale, denoted $m_D$, of $\delta_e$, $\delta_\mu$,
and $\delta_\tau$ --- assuming that these parameters are all
of the same order of magnitude.\footnote{$\delta_\mu$ and $\delta_\tau$
certainly are of similar magnitude,
as follows from Eqs.~(\ref{Mnu}),
(\ref{M'}),
and (\ref{rvalue}).}
\item The mass scale of the new scalars,
where we make the reasonable assumption 
$m_h \sim m_f \sim \langle \eta \rangle_0$.
\item The mass $M$ of the right-handed neutrino singlets.
\end{itemize}
Obviously,
the two mass-squared differences
$\Delta m^2_{\rm atm}$ and $\Delta m^2_\odot$
cannot determine all three mass scales.
However,
the two-loop expression in Eq.~(\ref{2-loop}) results in
\begin{equation}\label{eqa}
a \sim
\frac{1}{(16 \pi^2)^2}\, \left| \lambda f_\tau^2 h_{\tau \tau} \right|
\frac{m_\tau^2}{m_h},
\end{equation}
where we have assumed that all the couplings $h_{\ell \ell'}$
are of similar magnitude
(and also $|f_\mu| \approx |f_\tau|$)
and that,
therefore,
the two-$\tau$ contribution to Babu's two-loop diagram is dominant.
Equation~(\ref{eqa}) allows us to estimate
the mass scale of the Higgs scalars:
\begin{equation}\label{mh}
m_h \sim \left| \lambda f_\tau^2 h_{\tau \tau} \right|
\times 10^{14} \; \mbox{GeV}.
\end{equation}
With the dimensionless coupling constants $\lambda$,
$f_\tau$, and $h_{\tau \tau}$
being at most of order 1,
we may consider $10^{14}$ GeV as an upper bound for $m_h$.
Clearly,
with the scalar singlets having such large masses,
there are no restrictions on their Yukawa interactions
stemming from decays and scattering data. 

From the atmospheric mass-squared difference we get
\begin{equation}
\left| \frac{m_D^2}{M} \right| \sim 
\sqrt{\Delta m^2_{\rm atm}} \sim 0.06\; \mbox{eV}.
\end{equation}
We can tentatively fix $m_D$
by making the assumption that it is of the order of $m_\tau$,
the largest of the charged-lepton masses.
This leads to $M \sim 10^{11}$ GeV.
Interestingly,
this order of magnitude is compatible with $m_h$ ---
see Eq.~(\ref{mh}) ---
if we assume that the dimensionless coupling constants are $\sim 10^{-1}$.
An attractive
option would be to identify the two mass scales,
{\it i.e.},
to assume $M \sim m_h$.

To summarize,
in this paper we have advocated the three-parameter neutrino mass matrix
in Eq.~(\ref{M'}).
That mass matrix yields maximal solar-neutrino mixing and,
therefore,
it requires the vacuum-oscillation solution
to the solar-neutrino deficit
(the LOW solution might also be an option).
On the other hand,
in order to obtain nearly maximal atmospheric neutrino mixing
one has to tune the parameter $r$ to be close to 1. 
With this mass matrix one gets the neutrino mass spectrum
$m_1 \simeq m_2$ and $m_3 = 0$,
while $U_{e3}=0$ in the lepton mixing matrix of Eq.~(\ref{U}).
Notice that $m_3 = 0$, the inverted mass hierarchy, 
and $U_{e3} = 0$ are exact, testable predictions.
We have furthermore shown that
the mass matrix in Eq.~(\ref{M'}) can be obtained
in an extension of the Standard Model
with a spontaneously broken lepton-number symmetry $\bar L$
--- the ensuing Majoron couples very weakly to matter ---
together with a seesaw mechanism,
responsible for the smallness of the mass $b$,
and a radiative mass-generation mechanism,
responsible for the tiny mass $a$.
We have shown that it is sufficient to have
a single heavy mass scale in our model,
comprising both the seesaw scale
and the scale associated with the masses
of the new gauge-singlet scalar particles;
a value of order $10^{11}$ GeV would be a suitable choice for
this heavy mass scale.
The smallness of the mass $a$ in the mass matrix of Eq.~(\ref{M'})
practically forbids neutrinoless $\beta \beta$ decay.


\begin{thebibliography}{99}


\bibitem{SK-atm}
Super-Kamiokande Collaboration,
Y.\ Fukuda {\it et al.},
Phys.\ Rev.\ Lett.\ {\bf 81}, 1562 (1998); {\bf 82}, 2644 (1999);
Phys.\ Lett.\ B {\bf 467}, 185 (1999).

\bibitem{barr}
For recent reviews, see G.\ Altarelli and F.\ Feruglio,
Phys.\ Rep.\ (to be published),
hep-ph/9905536;
S.\ M.\ Barr and I.\ Dorsner,
hep-ph/0003058.

\bibitem{review}
For recent reviews, see
K.\ Zuber, Phys.\ Rep.\ {\bf 305}, 295 (1998);
S.\ M.\ Bilenky, C.\ Giunti, and W.\ Grimus,
Prog.\ Part.\ Nucl.\ Phys.\ {\bf 43}, 1 (1999);
P.\ Fisher, B.\ Kayser, and K.\ S.\ McFarland,
Ann.\ Rev.\ Nucl.\ Part.\ Sci.\ {\bf 49}, 
edited by C.\ Quigg, V.\ Luth, and P.\ Paul 
(Annual Reviews, Palo Alto, California, 1999), p.\ 481;
V.\ Barger, hep-ph/0003212.

\bibitem{seesaw}
M.\ Gell-Mann, P.\ Ramond, and R.\ Slansky,
in {\it Supergravity},
edited by D.\ Z.\ Freedman and F.\ van Nieuwenhuizen
(North Holland, Amsterdam, 1979);
T.\ Yanagida,
in {\it Proceedings of the workshop on unified theory and baryon 
number in the universe},
edited by O.\ Sawata and A.\ Sugamoto
(KEK, Tsukuba, Japan, 1979);
R.\ N.\ Mohapatra and G.\ Senjanovi\'c,
Phys.\ Rev.\ Lett.\ {\bf 44}, 912 (1980).

\bibitem{konetschny}
W.\ Konetschny and W.\ Kummer,
Phys.\ Lett.\ {\bf 70B}, 433 (1977).

\bibitem{zee}
A.\ Zee,
Phys.\ Lett.\ {\bf 93B}, 389 (1980); {\bf 161B}, 141 (1982).

\bibitem{jarlskog}
C.\ Jarlskog, M.\ Matsuda, S.\ Skadhauge, and M.\ Tanimoto,
Phys.\ Lett.\ B {\bf 449}, 240 (1999);
P.\ H.\ Frampton and S.\ L.\ Glashow, 
{\it ibid.} {\bf 461}, 95 (1999).

\bibitem{babu}
K.\ S.\ Babu,
Phys.\ Lett.\ B {\bf 203}, 132 (1988).

\bibitem{joshi}
A.\ S.\ Joshipura and S.\ D.\ Rindani, 
Phys. Lett. B {\bf 464}, 239 (1999).

\bibitem{luis}
L.\ Lavoura,
hep-ph/0005321.

\bibitem{kitabayashi}
T.\ Kitabayashi and M.\ Yasu\`e,
hep-ph/0006014.

\bibitem{CHOOZ}
CHOOZ Collaboration, M.\ Apollonio {\it et al.},
Phys.\ Lett.\ B {\bf 466}, 415 (1999).

\bibitem{Ue3}
O.\ Yasuda,
Phys.\ Rev.\ D {\bf 58}, 091301 (1998);
G.\ L.\ Fogli {\it et al.},
{\it ibid.} {\bf 59}, 033001 (1999);
V.\ Barger and K.\ Whisnant,
{\it ibid.} {\bf 59}, 093007 (1999).

\bibitem{barbieri}
S.\ T.\ Petcov,
Phys.\ Lett.\ {\bf 110B}, 245 (1982);
C.\ N.\ Leung and S.\ T.\ Petcov,
{\it ibid.} {\bf 125B}, 461 (1983); 
R.\ Barbieri, L.\ J.\ Hall, D.\ Smith, A.\ Strumia, and N.\ Weiner,
JHEP {\bf 9812}, 017 (1998).

\bibitem{hooft}
G.\ 't Hooft, 
in {\it Recent developments in gauge theories, Carg\`ese 1979},
edited by G.\ 't Hooft {\it et al.}
(Plenum Press, New York, 1980).

\bibitem{majoron}
Y.\ Chikashige, R.\ N.\ Mohapatra, and R.\ D.\ Peccei, 
Phys.\ Lett.\ {\bf 98B}, 265 (1981).

\bibitem{branco}
G.\ C.\ Branco, W.\ Grimus, and L.\ Lavoura, 
Nucl.\ Phys.\ {\bf B312}, 492 (1989). 

\bibitem{VO}
J.\ N.\ Bahcall, P.\ I.\ Krastev, and A.\ Yu.\ Smirnov,
Phys.\ Rev.\ D {\bf 58}, 096016 (1998); {\bf 60}, 093001 (1999).

\bibitem{barger} 
V.\ Barger and K.\ Whisnant,
Phys.\ Lett.\ B {\bf 456}, 54 (1999).

\bibitem{goswami}
S.\ Goswami, D.\ Majumdar, and A.\ Raychaudhuri,
hep-ph/9909453.

\bibitem{quasi}
G.\ L.\ Fogli, E.\ Lisi, D.\ Montanino, and A.\ Palazzo,
hep-ph/0005261.

\bibitem{GG}
M.\ C.\ Gonzalez-Garcia {\it et al.}, 
Nucl.\ Phys.\ {\bf B573}, 3 (2000).

\bibitem{dark}
M.\ C.\ Gonzalez-Garcia and C.\ Pe\~na-Garay,
hep-ph/0002186.

\bibitem{giunti}
S.\ M.\ Bilenky and C.\ Giunti, 
Phys.\ Lett.\ B {\bf 444}, 379 (1998).

\bibitem{sobel}
Super-Kamiokande
Collaboration,
H.\ Sobel,
talk presented at Neutrino2000,
June 16--21, 2000, Sudbury, Ontario, in
http://nu2000.sno.laurentian.ca/H.Sobel/.
See also N.\ Fornengo, M.\ C.\ Gonzalez-Garcia, and J.\ W.\ F.\ Valle,
hep-ph/0002147.

\bibitem{bimaximal}
E.\ Torrente-Lujan,
Phys.\ Lett.\ B {\bf 389}, 557 (1996);
F.\ Vissani,
hep-ph/9708483;
V.\ Barger, S.\ Pakvasa, T.\ J.\ Weiler, and K.\ Whisnant,
Phys.\ Lett.\ B {\bf 437}, 107 (1998);
M.\ Je\.zabek and Y.\ Sumino, 
{\it ibid.} {\bf 440}, 327 (1998);
A.\ Baltz, A.\ S.\ Goldhaber, and M.\ Goldhaber, 
Phys.\ Rev.\ Lett.\ {\bf 81}, 5730 (1998).

\end{thebibliography}
\end{document}